\documentclass[reviewcopy]{elsart}
\usepackage{rotating,epsfig}

\usepackage{psfig}

\begin{document}

\begin{frontmatter}

\title{
Quantifying Influenza Vaccine Efficacy and Antigenic Distance
}

\author{Vishal Gupta, David J.\ Earl, and Michael W.\ Deem}
\address{
Departments of Bioengineering and Physics \& Astronomy\\
Rice University\\
6100 Main Street---MS 142\\
Houston, TX\ \ 77005--1892
}


\bigskip

\address{M.\ W.\ Deem, mwdeem@rice.edu,
tel: 713--348--5852, fax: 713--348--5811}

\begin{abstract}
We introduce a new measure of antigenic distance between
influenza A vaccine and circulating strains.
The measure correlates well with efficacies of the H3N2 influenza
A component of the annual vaccine between 1971 and 2004,
as do results of a theory of the immune
response to influenza following vaccination.
This new measure of antigenic distance is correlated 
with vaccine efficacy to a greater degree 
than are current state-of-the-art
phylogenetic sequence analyzes or 
ferret antisera inhibition assays.
We suggest that this new measure
of antigenic distance be used in the design of the
annual influenza vaccine and in the interpretation of
vaccine efficacy monitoring.
\end{abstract}
\end{frontmatter}

\section{Introduction}

Annual influenza epidemics are responsible for the deaths of 250000 to
500000 people worldwide and cause illness in 5 to 15\% of the 
total population each year \cite{WHO_2}. The total direct and indirect
costs associated with influenza in the USA
are roughly \$10 billion \cite{Lave}, and 
the economic cost of an influenza pandemic is estimated
to be between \$71--167 billion \cite{Meltzer} in the USA alone.
Vaccination is the primary method employed to prevent infection by influenza 
and its associated complications.  Antigenic change, combined with
the high transmission rate of influenza strains, means that the vaccine
must be redesigned annually, currently based upon phylogenetic, 
 experimental, and epidemiological analysis.

The effectiveness of the annual influenza vaccine varies from year
to year due to changes in the identity of the circulating influenza
strains.  Typically, three strains are included in the annual vaccine,
with these three strains chosen to be as similar as possible to
those projected to be
the most prominent circulating strains in the upcoming influenza season.
Currently, the vaccine contains H3N2 and H1N1 influenza A components and
an influenza B component.
Since the mutation rate of the
influenza virus is rather high, vaccine efficacies are rarely 
100\%, and are more typically 30 -- 60\%, against influenza-like illness.
As significant as the estimated worldwide mortality is,
it rises by another factor of 160\% \cite{Neuzil} to 
260\% \cite{Sprenger}
if influenza-induced complications to patients with
other conditions are included, and the influenza vaccine on average
significantly reduces such excess mortality \cite{Hak}.
Vaccine efficacy
can even be negative, however,
due to original antigenic sin \cite{Davenport,Fazekas1,Deem},
the tendency for antibodies produced in response to exposure to
influenza vaccine antigens to
suppress the creation of new, different antibodies in response to exposure to
new versions of the influenza virus.
The efficacy of the annual influenza
vaccine, and whether original antigenic sin
may occur, depends sensitively on how similar the vaccine and 
circulating viral strains are. 
Current state of the art measures of antigenic distance are
based on ferret antisera hemagglutinin inhibition 
assays \cite{Smith2,Smith,sera},
and these distances are assumed to correlate well with vaccine
efficacies in humans. However, to our knowledge no such good
correlation has ever been shown for an experimental or theoretical 
measure of antigenic distance.
In addition to its annual value,
a reliable measure of antigenic distance would provide
valuable extra time if preparation and rush production of a 
modified vaccine is necessary
to stem the spread of a newly emerged influenza strain \cite{Ault}.
We here provide a quantitative definition of
the difference between dominant epitope regions
in the vaccine and circulating strain and show that
this definition of antigenic distance correlates well
with human influenza vaccine efficacy over the last
35 years.

\section{Methods}

\begin{figure}[t]
\begin{center}
\epsfig{file=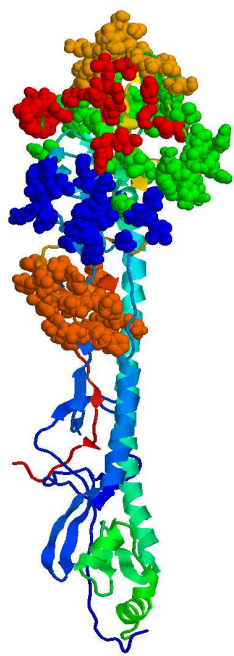,height=5in,clip=,angle=-90}
\end{center}
\caption{Hemagglutinin protein for the A/Fujian/411/2002 strain.
Highlighted are the A (red), B (orange), C (brown), D (green), and
E (blue) epitopes \cite{ISD}. 
The rest of the protein is shown in ribbon format.
\label{fig:HA}}
\end{figure}

Our theory models the response of an immune system not subject
to immunosenescence. Therefore, we limited consideration to
experimental studies of
vaccine efficacy for 18 to 64 year old subjects in all years 
since sequencing began, when
the H3N2 subtype of influenza A was
the predominant virus,
and where epidemiological data on vaccine efficacy exists.
We focus on the H3N2 strain because
it is the most common strain and is responsible for significant
morbidity and mortality and due
to the abundance of available crystallographic, genetic, and epidemiological
data.  
Our approach, however, is general.  
As is customary, we restrict attention to the hemagglutinin protein,
against which neutralizing antibodies are generated \cite{Bush99}.
Shown in figure \ref{fig:HA} is the hemagglutinin protein for the 
A/Fujian/411/2002 strain with the epitope regions highlighted.

Our theory of the immune response to vaccination and disease uses the
generalized $NK$ model \cite{Deem}
 to calculate
affinity constant
values for the immune response to an antigen following vaccination.
In this theory, the natural order parameter to distinguish
between the vaccine strain and the circulating strain
is the fraction of amino acids that differ
in the dominant epitope region. The model considers the
diversity of an individual's antibody repertoire and includes interactions
within the antibody and between the antibody and the antigen.
Here, the binding constant is determined as $K = \exp(a - bU)$, where
$a=-18.56$ and $b=1.67$ are found from a comparison of the dynamics
of the model with experiment \cite{Deem}, and $U$ is the energy function
for an antibody \cite{Deem}.

To apply the theory to a candidate vaccine and circulating strain,
the sequences and identity of the dominant epitope must be known.
The sequences
and identities of the vaccine and circulating strains for each year
were taken from Ref.\ \cite{ISD}.
The definition of the five epitopes, or surface regions that
are recognized by human antibodies, in the H3N2 hemagglutinin protein
were also taken from Ref.\ \cite{ISD}.
The dominant epitope, or the epitope that induces the most significant
immune response, for a particular circulating strain in
a particular year was taken
as that which had the largest fractional change in amino acid
sequence relative to the vaccine
strain \cite{Fitch,Bush,Plotkin,Fitch2}.
We use as our definition of antigenic distance  $p_{\rm epitope}$,
where
\begin{equation}
 p_{\rm epitope} =
    \frac{{\rm number\: of\: amino\: acid\: differences\: in\:
the\: dominant\: epitope}}
   {{\rm total\: number\: of\: amino\: acids\: in\: the\:
dominant\: epitope}} \ .
\label{eq_epitope}
\end{equation}

The difference
between the vaccine strain and the circulating strain is
defined in the model by $p_{\rm epitope}$, eq.\ \ref{eq_epitope}.
The vaccine efficacy, $E$, was assumed to correlate with the
binding constant as $E = \alpha \ln [ K_{\rm secondary}(p_{\rm epitope})
 / K_{\rm primary}]$, where the constant $\alpha$ is chosen so that
a perfect match between the vaccine and circulating strain
leads to 47\% vaccine efficacy (which matches
historical data, see table 1),
$K_{\rm primary}$ is
the binding constant for the primary immune response, and $K_{\rm secondary}$
is the binding constant for the secondary immune response following
vaccination.  The theory is entirely predictive, with no
fitted parameters save for the determined constant $\alpha$.
For example, the point at which the vaccine efficacy becomes
negative is independent of the value of $\alpha$.

\section{Results}

Shown in figure \ref{fig:epitope} and table 1 are the experimental 
vaccine efficacies and the efficacies 
predicted by the theory as a function of $p_{\rm epitope}$. 
Vaccine efficacies are taken from the 
literature \cite{Vish32,Vish41,Vish10,Vish16,Vish35,Vish33,Vish1,Vish31,Vish37,Vish7,Vish43,VishFrance,VishJap,VishCan,CDC} 
and defined as
\begin{equation}
(u - v) / u \ , 
\label{eq_efficacy}
\end{equation}
where $u$ is the influenza-like illness rate of unvaccinated 
individuals, and $v$ is the influenza-like illness rate of vaccinated 
individuals. While the epidemiological estimates of $u$ and $v$
contain statistical noise, these are the best estimates available
of vaccine efficacy in humans.
The statistical mechanical model captures the essential physics of the immune
response to influenza vaccination and demonstrates the value of using
$p_{\rm epitope}$ to define the degree of
antigenic drift.
Consideration of antigenic
drift of the dominant epitope follows
from immunoassays and crystallographic
data that show only the epitope regions are significantly involved in
immune recognition \cite{Epitope2}.
When the antigenic drift, $p_{\rm epitope}$, in the dominant epitope
is greater than 0.19, according to historical records, or 0.22,
according to theory, the vaccine efficacy becomes negative
(see figure \ref{fig:epitope}).
This regime is to be avoided.
For example, in the 1997/1998 northern hemisphere influenza season, when the 
Sydney/5/97 strain became widespread, $p_{\rm epitope} = 0.238$, 
and the vaccine
efficacy was -18 \% \cite{Vish7}.
The only data point that falls significantly off the
theory is that for the 1989/1990 epidemic \cite{1762}, in which
it is likely that multiple circulating strains were present,
including influenza B strains \cite{infB,Vish35}.


When the vaccine efficacy is compared to the sequence difference of
the entire hemagglutinin protein,
\begin{equation}
 p_{\rm sequence} =
    \frac{{\rm number\: of\: amino\: acid\: differences\: in\:
the\: sequence}}
   {{\rm total\: number\: of\: amino\: acids\: in\: the\: 
sequence}} \ ,
\label{eq_sequence}
\end{equation}
one current measure of antigenic drift
used to construct phylogenetic relationships between circulating
strains for the WHO February report \cite{WHO},
the correlation is far less apparent.
These data are shown in  figure \ref{fig:sequence} and table 1.  
Since much of the
protein is inaccessible to antibodies or simply not recognized by
human antibodies, drift in  much of the protein sequence is not
correlated with vaccine efficacy.


When the vaccine efficacy is compared to the antigenic distance
derived from ferret antisera \cite{Smith,sera},
 the dominant current measure of antigenic drift
used to confirm phylogenetic strain analysis \cite{WHO},
the correlation is again less apparent.
These data are shown in  figure \ref{fig:ferret} and table 1.  It appears
that the ferret antisera experiments capture no more information
than does the analysis with $p_{\rm sequence}$.
A ferret-derived antigenic distance of zero does not
always guarantee that the two strains are identical.  For example,
for the 1996/1997 vaccine strain of A/Nanchang/933/95
and circulating strain of A/Wuhan/359/95, the
ferret-derived antigenic distance was zero, whereas, $p_{\rm epitope} = 0.095$,
and the vaccine efficacy was 27\% in the northern
hemisphere \cite{VishFrance}
and 11\% the next year in the southern hemisphere \cite{Vish43}.
These values are
much lower than the average for a perfect match between
vaccine and circulating strains, which  is 47\%.

\section{Discussion}

Vaccine design is done under considerable time pressure.
At present, the WHO and national health agencies in the northern
hemisphere determine the components of the annual flu
vaccine between February and April. The vaccine is then produced by
growing virus in hen's eggs, and it is distributed in September after 
regulatory tests in mid-July \cite{1776}.
Data collection relating to the effectiveness of the vaccine can begin in 
October, and by January a very good measure of the season's vaccine
has been obtained.  The availability of high-growth
reassortments from egg-cultured strains imposes additional
constraints on the choice of possible vaccine strains.
Given the constraints imposed by the biology and manufacturing
process, one wishes to choose the strain that provides the best possible
match to the anticipated circulating strain for the following season.

\begin{figure}[t]
\begin{center}
\epsfig{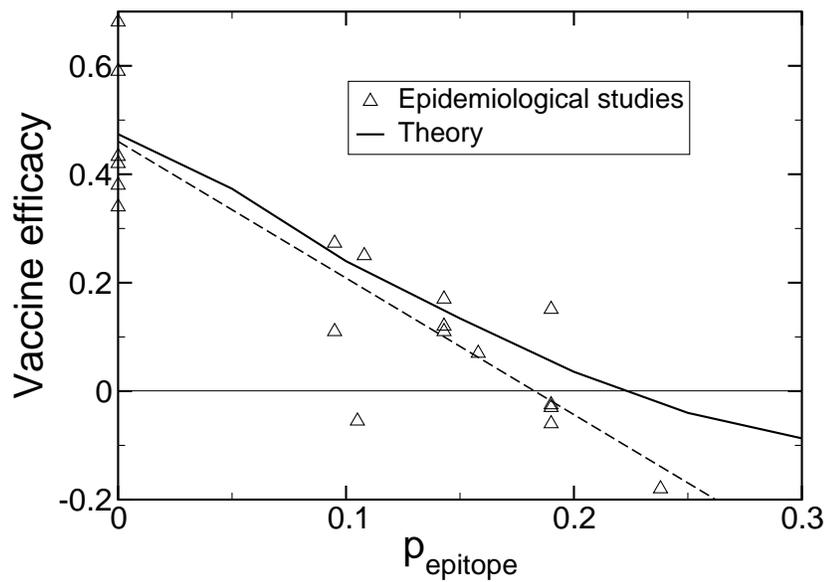}
\end{center}
\caption{Vaccine efficacy for influenza-like illness as a function
of $p_{\rm epitope}$ as observed in epidemiological studies and as
predicted by theory.  Also shown is a linear least squares fit to
the data (long dashed, $R^2 = 0.79$).
\hspace{3in}
\label{fig:epitope}}
\end{figure}
\begin{figure}[t]
\begin{center}
\epsfig{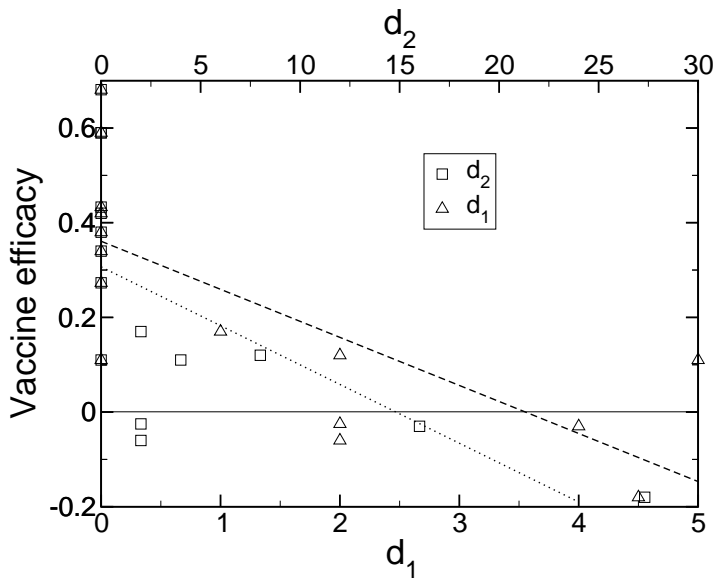}
\end{center}
\caption{Vaccine efficacy for influenza-like illness as a function of
two measures of antigenic distance, $d_1$ \cite{Smith} and $d_2$ \cite{sera},
derived from ferret antisera experiments.
Experimental data were collected from a variety of 
sources \cite{Smith,WHO,sera2,sera3,sera4,sera5,sera7,CDC_97}.
Results were averaged when
multiple hemagglutination inhibition (HI) studies had 
been performed for a given year. These HI binding assays measure the ability
of ferret antisera to block the agglutination of red blood cells by 
influenza viruses. Also shown are linear least squares fits
to the $d_1$ (long dashed, $R^2 = 0.54$) and $d_2$ (short dashed,
$R^2 = 0.41$) data.
The epidemiological data shown in this figure  are the same 
as in figure \ref{fig:epitope}.  Only the definition of the 
$x$-axis is different.
\hspace{3in}
\label{fig:ferret}}
\end{figure}

We believe that the 
antigenic distance between strains would profitably be defined by
world health professionals as $p_{\rm epitope}$ (figure \ref{fig:epitope}), 
rather than as sequence distance (figure \ref{fig:sequence}) or from 
ferret antisera assays (figure \ref{fig:ferret}).  
Of importance to note is that the immune response is non-monotonic and
non-linear in the antigenic distance, \emph{i.e.}\ original antigenic 
sin or negative vaccine efficacy exists only for an intermediate antigenic 
distance. In this regime, the vaccine can induce a greater degree 
of susceptibility to flu-like illness in vaccinated individuals relative to
unvaccinated individuals.  This negative efficacy has occurred
26\% of the time for circulating H3N2 strains in the last 33 years
(5 of the 19 data points in table 1, figure \ref{fig:epitope} and figure 
\ref{fig:sequence} are negative).
Thus, original antigenic sin can occur not only
if an individual's flu shot is not updated
on an annual basis, but also even if an individual's flu shot
is updated yearly.
The original antigenic sin regime is to be avoided both for the
immunological consequences and for the negative impact of such
a vaccine on public health policy acceptance.  Our theory quantifies
where the regime lies and lends additional credence to the experimental
measurements of such negative vaccine efficacies.
While negative efficacies have often been thought to be
experimental error (and appear to be noise in figure \ref{fig:sequence}),
they are not.  Negative efficacies appear only for large
values of $p_{\rm epitope}$ (see figure \ref{fig:epitope}).


As an example of how our theory can be used to help guide
public health policy, we apply it to the 2004/2005 northern
hemisphere flu season.
By using $p_{\rm epitope}$ as the definition of antigenic distance,
one may be quantitative about which strains will \emph{a priori}
be most protective, and so should be chosen
for inclusion in the annual vaccine.
For example,
to combat the A/Fujian/411/2002 strain that was predominant in the
2003--2004 influenza epidemic, the CDC council decided to use
A/Wyoming/3/2003, a strain termed `antigenically equivalent' to
A/Fujian/411/2002, as the
H3N2 component of the 2004--2005 vaccine \cite{MMWR04}.
Our analysis yields
$p_{\rm epitope} = 0.095$ between these two strains, suggesting that
the vaccine will have an efficacy of roughly 20\%
for influenza-like illness
against the Fujian strain (see figure \ref{fig:epitope}),
and that these strains are not antigenically equivalent.
Conversely, for A/Kumamoto/102/2002, another
available H3N2 component \cite{WHO_weekly}, we find
$p_{\rm epitope} = 0$ versus A/Fujian/411/2002,
suggesting this component would provide superior protection to Fujian
than would the Wyoming strain.

\begin{figure}[t]
\begin{center}
\epsfig{file=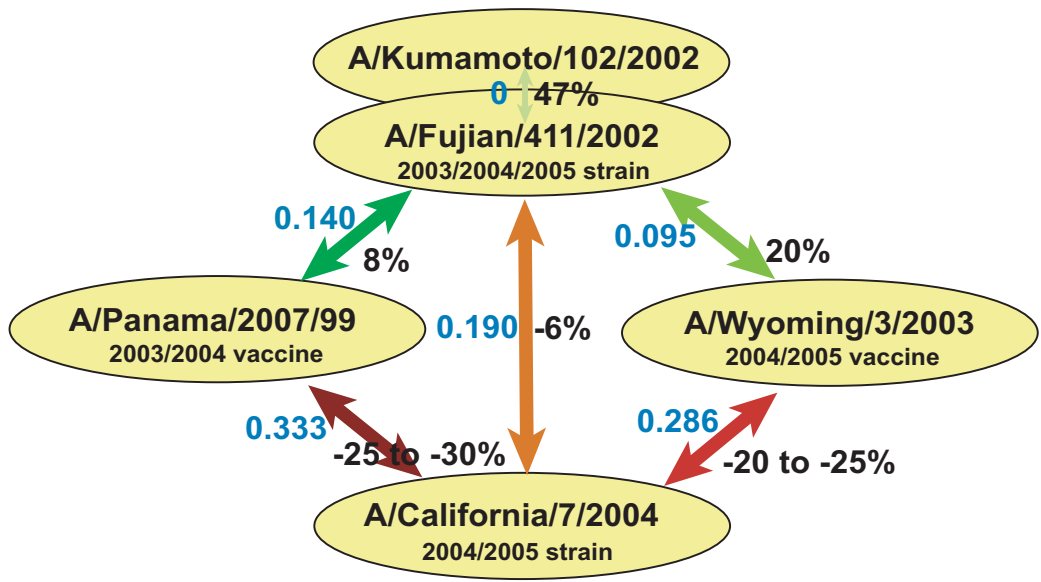,height=3in,clip=,angle=0}
\end{center}
\caption{$p_{\rm epitope}$ (blue) and estimated vaccine efficacy
(black) from figure \ref{fig:epitope} between components used in the
2003/2004 and 2004/2005 vaccinations and circulating strains in the
2004/2005 season.
\hspace{3in}
\label{fig:strains}}
\end{figure}

Continuing this example of how our theory can be used in
vaccine design, we show in figure \ref{fig:strains} the calculated
$p_{\rm epitope}$ values and vaccine efficacies between recent
influenza A H3N2 vaccine components and circulating strains.
Many isolates from the 2004/2005 flu season have been
A/Fujian/411/2002-like strains \cite{WHO_weekly2}.
Another circulating strain that began to emerge in late 2004
is
A/California/7/2004, and an A/California/7/2004-like strain is
recommended as the influenza A component of the 2005/2006 northern
hemisphere vaccine by the WHO \cite{WHO_2005north},
suggesting that this is an important strain to consider as an example.
For individuals
who received a vaccination in 2003/2004 (the A/Panama/2007/99 strain) and
who were not exposed to the Fujian strain, their protection against
the Fujian strain is low, and their protection against
the California strain is
in the region of original antigenic sin.
For individuals who were vaccinated in the 2004/2005 season
(the A/Wyoming/3/2003 strain), their protection against the Fujian
strain is moderate, but their protection against
the California strain is again in the region
 of original antigenic sin.
For individuals who were exposed to the Fujian strain in 2003/2004 or
2004/2005, their protection against
the California strain is just in the region of original antigenic sin.
The 2005 southern hemisphere vaccine strain was A/Wellington/1/2004.
Our analysis yields $p_{\rm epitope}=0.143$ between Wellington and 
California indicating the vaccine will provide a low level 
of protection against the California strain.
These findings suggest that production of a new vaccine strain
to combat A/California/7/2004 in the next flu season is essential.
Persons who recieved a flu vaccine in 2003/2004 and/or 2004/2005
should be particularly encouraged to receive a flu shot in 
2005/2006 as they are likely to be more susceptible to this new strain
than if they had not received their flu shot in the previous 2 years.


We believe that greater resources need to be applied to
the experimental study of the efficacy of the flu vaccine each year.
For example, in order to calculate the antigenic distance
optimally, the dominant human epitope in each strain is needed,
and is not currently measured.  Measurement of which epitope
is dominant for each vaccine and circulating strain should increase the
predictive ability of our approach, beyond that
in figure \ref{fig:epitope}.
More epidemiological studies relating
antigenic drift to vaccine efficacy are needed \cite{CDC_call}
and would help guide the
management of health resources during the flu season.
Since substantial
costs are associated with lost work due to influenza among those
in the 18-64 age bracket, large studies of this age range are both important
and informative, due to lack of immunosenescence. 
Continuous measurement and sequencing 
of the dominant circulating strains during
the flu season, combined with the theory of figure \ref{fig:epitope},
should enable better prediction of the severity of the annual flu season
and better design of the subsequent year's vaccine.

More generally, our results have implications for the design of 
vaccines to combat rapidly mutating
viral diseases that are controlled by
antibody responses.  We suggest that antigenic drift in the
dominant epitope, $p_{\rm epitope}$, will provide a 
prediction measure of efficacy for such vaccines.
This quantitative measure of efficacy
may then be used to determine the frequency and 
nature of vaccine redesign that is necessary.

\section*{\bf Acknowledgments}
This research was supported by the U.S.\ National Institutes of Health vaccine
group and the National Science Foundation.

\hbox{}\hspace{-\parindent}{\bf Competing Financial Interests}
Authors declare no competing financial interests.

\hbox{}\hspace{-\parindent}{\bf Correspondence}
Correspondence should be addressed to
MWD (mwdeem@rice.edu).
\\

\section*{Appendix: The Generalized $NK$ Model}

Our theory of the immune response to vaccination and disease uses the 
generalized $NK$ model \cite{Deem}
 to calculate 
affinity constant
values for the immune response to an antigen following vaccination.
In this theory, the natural order parameter to distinguish
between the vaccine strain and the circulating strain 
is the fraction of amino acids that differ
in the dominant epitope region, $p_{\rm epitope}$. The model considers the
diversity of an individual's antibody repertoire and includes interactions
within the antibody and between the antibody and the antigen.
Here, the binding constant is determined as 
$K = \exp(a - b \langle U \rangle)$, where
$a=-18.56$ and $b=1.67$ are found from a comparison of the dynamics
of the model with experiment \cite{Deem} and $U$ is the energy function
for an antibody and is defined as
\begin{equation}
 U = \sum_{i = 1}^M U_{\alpha_i}^\mathrm{sd} +
\sum_{i > j = 1}^M U_{ i j }^\mathrm{sd-sd}
+ \sum_{i = 1}^P U_i^\mathrm{c} \ .
\label{101}
\end{equation}
The parameters within the
generalized block $NK$ model represent the number of secondary
structures and the total size of the variable region \cite{Deem}.
We have $L=5$ different subdomain energy functions of the $NK$ form
\begin{equation}
U_{\alpha_{i}}^\mathrm{sd} = \frac{1}{\left[ M(N-K)\right]^{1/2}}
\sum_{j = 1}^{N-K+1} \sigma_{\alpha_{i}} \left(
a_j, a_{j+1}, \ldots, a_{j+K-1}
\right) \ ,
\label{102}
\end{equation}
where $a_{j}$ is the amino acid type of the $j$th amino acid in the
subdomain, and $\alpha_i$ is the type of the $i$th subdomain.
 As in previous studies, we consider the case where the
range of the interactions within a subdomain is specified by
$K=4$ and there are $N=10$ amino acids in
each subdomain \cite{Bogarad_Deem}.
Here $\sigma_{\alpha_{i}}$ is a quenched Gaussian random number with zero
mean and a variance of unity, and it is different for each value of its
argument for each of the $L$ subdomain types, $\alpha_{i}$. The
interaction energy between secondary subdomain structures is given by
\begin{eqnarray}
U_{i j }^\mathrm{sd-sd} &=& \left[
\frac{2}{D M (M-1)} \right]^{1/2} \nonumber \\
&&\times
\sum_{k=1}^D \sigma_{i j }^{(k)}
\left(
a_{j_1}^{(i)} , \dots,
a_{j_{K/2}}^{(i)};
a_{j_{K/2+1}}^{(j)}, \ldots
a_{j_{K}}^{(j)}
 \right) \ .
\label{103}
\end{eqnarray}
Here $M=10$ is the number of antibody secondary structural subdomains.
We consider $D=6$ interactions between secondary
structures \cite{Bogarad_Deem}.
The zero-mean, unit-variance Gaussian $\sigma_{ i j }^{(k)}$ and
the interacting amino acids,
$j_{1}, \ldots, j_{K}$, are selected at random for each interaction
($i, j, k $). In our model, $P=5$ amino acids contribute directly
to an antigen binding event, where the chemical binding energy
of each amino acid is given by
\begin{equation}
U_i^\mathrm{c} = \frac{1}{\sqrt P} \sigma_i \left( a_i \right) \ ,
\label{104}
\end{equation}
where the zero-mean, unit-variance Gaussian
$\sigma_i$ and the contributing amino acid, $i$, are chosen at random.

To model the immune system dynamics, we use 30 rounds of point
mutation and selection to evolve
our antibody sequences, which corresponds
to an immune response of approximately 10 days.
For each round of selection we conduct 0.5 point mutations per antibody
sequence and amplify the best 20\% of antibody sequences to form
the starting population for the next round of selection.
The secondary immune
response following vaccination uses evolved memory sequences as well
as naive cells, whereas the primary immune response uses
only naive cells \cite{Deem}.

\bibliography{influenza}

\clearpage

{
\renewcommand{\baselinestretch}{1.0} \tiny\normalsize

\begin{sidewaystable}
\centering
\begin{tabular}{l|c|c|c|c|c|c|c|c}
{\tiny Year}&
{\tiny Vaccine}&
{\tiny Circulating}&
{\tiny Vaccine}&
{\tiny Dominant}&
{\tiny $p_{epitope}$}&
{\tiny $p_{sequence}$}&
{\tiny $d_1$}&
{\tiny $d_2$}\\
&
{\tiny Strain}&
{\tiny Strain}&
{\tiny Efficacy}&
{\tiny Epitope}&
&
&
&
\\
\hline
\hline
{\tiny 1971-72}&
{\tiny Aichi/2/68 (V01085)}&
{\tiny HongKong/1/68 (AF201874)}&
{\tiny $7$ \% \cite{Vish32}}&
{\tiny A}&
{\tiny 0.158}&
{\tiny 0.033}&
\\
{\tiny 1972-73}&
{\tiny Aichi/2/68 (V01085)}&
{\tiny England/42/72 (AF201875)}&
{\tiny $15$ \% \cite{Vish32}}&
{\tiny B}&
{\tiny 0.190}&
{\tiny 0.055}&
{\tiny }&
{\tiny }
\\
{\tiny 1973-74}&
{\tiny England/42/72 (ISDNENG72)}&
{\tiny PortChalmers/1/73 (AF092062)}&
{\tiny $11$ \% \cite{Vish32}}&
{\tiny B}&
{\tiny 0.143}&
{\tiny 0.018}&
{\tiny 5 \cite{Smith}}&
{\tiny 4 \cite{Smith}}
\\
{\tiny 1975-76}&
{\tiny PortChalmers/1/73 (AF092062)}&
{\tiny Victoria/3/75 (ISDNVIC75)}&
{\tiny $-3$ \% \cite{Vish32}}&
{\tiny B}&
{\tiny 0.190}&
{\tiny 0.055}&
{\tiny 4 \cite{sera2}}&
{\tiny 16 \cite{sera2}}
\\
{\tiny 1984-85}&
{\tiny Philippines/2/82 (AF233691)}&
{\tiny Mississippi/1/85 (AF008893)}&
{\tiny $-6$ \% \cite{Vish41}}&
{\tiny B}&
{\tiny 0.190}&
{\tiny 0.033}&
{\tiny 2 \cite{sera5}}&
{\tiny 2 \cite{sera5}}
\\
{\tiny 1985-86}&
{\tiny Philippines/2/82 (AF233691)}&
{\tiny Mississippi/1/85 (AF008893)}&
{\tiny $-3$ \% \cite{Vish10,Vish16}}&
{\tiny B}&
{\tiny 0.190}&
{\tiny 0.033}&
{\tiny 2 \cite{sera5}}&
{\tiny 2 \cite{sera5}}
\\
{\tiny 1987-88}&
{\tiny Leningrad/360/86 (AF008903)}&
{\tiny Shanghai/11/87 (AF008886)}&
{\tiny $17$ \% \cite{Vish10,Vish35}}&
{\tiny B}&
{\tiny 0.143}&
{\tiny 0.024}&
{\tiny 2 \cite{sera5}}&
{\tiny 1 \cite{sera5}}
\\
{\tiny 1989-90}&
{\tiny Shanghai/11/87 (AF008886)}&
{\tiny England/427/88 (AF204238)}&
{\tiny $-6$ \% \cite{Vish35}}&
{\tiny A}&
{\tiny 0.105}&
{\tiny 0.021}&
{\tiny }&
{\tiny }
\\
{\tiny 1992-93}&
{\tiny Beijing/32/92 (AF008812)}&
{\tiny Beijing/32/92 (AF008812)}&
{\tiny $59$ \% \cite{Vish33}}&
{\tiny }&
{\tiny 0.0}&
{\tiny 0.0}&
{\tiny 0 \cite{sera3}}&
{\tiny 0 \cite{sera3}}
\\
{\tiny 1993-94}&
{\tiny Beijing/32/92 (AF008812)}&
{\tiny Beijing/32/92 (AF008812)}&
{\tiny $38$ \% \cite{Vish16}}&
{\tiny }&
{\tiny 0.0}&
{\tiny 0.0}&
{\tiny 0 \cite{sera3}}&
{\tiny 0 \cite{sera3}}
\\
{\tiny 1994-95}&
{\tiny Shangdong/9/93 (Z46417)}&
{\tiny Johannesburg/33/94 (AF008774)}&
{\tiny $25$ \% \cite{Vish1}}&
{\tiny A}&
{\tiny 0.105}&
{\tiny 0.021}&
{\tiny }&
{\tiny }
\\
{\tiny 1995-96}&
{\tiny Johannesburg/33/94 (AF008774)}&
{\tiny Johannesburg/33/94 (AF008774)}&
{\tiny $42$ \% \cite{Vish31}}&
{\tiny }&
{\tiny 0.0}&
{\tiny 0.0}&
{\tiny 0 \cite{sera4,CDC_97}}&
{\tiny 0 \cite{sera4,CDC_97}}
\\
{\tiny 1996-97}&
{\tiny Nanchang/933/95 (AF008725)}&
{\tiny Wuhan/359/95 (AF008722)}&
{\tiny $27$ \% \cite{VishFrance}}&
{\tiny B}&
{\tiny 0.095}&
{\tiny 0.006}&
{\tiny 0 \cite{sera4,CDC_97}}&
{\tiny 0 \cite{sera4,CDC_97}}
\\
{\tiny 1997}&
{\tiny Nanchang/933/95 (AF008725)}&
{\tiny Wuhan/359/95 (AF008722)}&
{\tiny $11$ \% \cite{Vish43}}&
{\tiny B}&
{\tiny 0.095}&
{\tiny 0.006}&
{\tiny 0 \cite{sera4,CDC_97}}&
{\tiny 0 \cite{sera4,CDC_97}}
\\
{\tiny 1997-98}&
{\tiny Nanchang/933/95 (AF008725)}&
{\tiny Sydney/5/97 (AJ311466)}&
{\tiny $-18$ \% \cite{Vish7}}&
{\tiny B}&
{\tiny 0.238}&
{\tiny 0.043}&
{\tiny 4.5 \cite{sera4,sera7}}&
{\tiny 27.3 \cite{sera4,sera7}}
\\
{\tiny 1998-99}&
{\tiny Sydney/5/97 (AJ311466)}&
{\tiny Sydney/5/97 (AJ311466)}&
{\tiny $34$ \% \cite{Vish7}}&
{\tiny }&
{\tiny 0.0}&
{\tiny 0.0}&
{\tiny 0 \cite{WHO,sera4}}&
{\tiny 0 \cite{WHO,sera4}}
\\
{\tiny 1999-00}&
{\tiny Sydney/5/97 (AJ311466)}&
{\tiny Sydney/5/97 (AJ311466)}&
{\tiny $43$ \% \cite{VishCan}}&
{\tiny }&
{\tiny 0.0}&
{\tiny 0.0}&
{\tiny 0 \cite{WHO,sera4}}&
{\tiny 0 \cite{WHO,sera4}}
\\
{\tiny 2001-02}&
{\tiny Panama/2007/99 (ISDNCDA001)}&
{\tiny Panama/2007/99 (ISDNCDA001)}&
{\tiny $68$ \% \cite{VishJap}}&
{\tiny }&
{\tiny 0.0}&
{\tiny 0.0}&
{\tiny 0 \cite{WHO,sera4}}&
{\tiny 0 \cite{WHO,sera4}}
\\
{\tiny 2003-04}&
{\tiny Panama/2007/99 (ISDNCDA001)}&
{\tiny Fujian/411/2002 (ISDN38157)}&
{\tiny $12$ \% \cite{CDC}}&
{\tiny B}&
{\tiny 0.143}&
{\tiny 0.040}&
{\tiny 2 \cite{WHO}}&
{\tiny 8 \cite{WHO}}
\\

\end{tabular}
\caption{Summary of results. The table includes the identities 
and accession numbers of the 
vaccine and circulating strains for each of the years since 1971 that
the H3N2 virus has been the predominant influenza virus and for
which vaccine efficacy data are available. Vaccine 
efficacy is taken from the 
literature \cite{Vish32,Vish41,Vish10,Vish16,Vish35,Vish33,Vish1,Vish31,Vish37,Vish7,Vish43,VishFrance,VishJap,VishCan,CDC} and
defined by eq.\ \ref{eq_efficacy}. Where
more than one study exists for the same influenza season,
the efficacy results are averaged.
 The dominant epitope is predicted by our theory for all
seasons where the vaccine and circulating strains are not a match.
The $p_{\rm epitope}$ and $p_{\rm sequence}$ 
values are calculated using eq.\  \ref{eq_epitope}
and eq.\ \ref{eq_sequence}, respectively. Two measures of antigenic distance
from ferret antisera assays, $d_1$ \cite{Smith}
and $d_2$ \cite{sera}, are determined
from the literature \cite{Smith,WHO,sera2,sera3,sera4,sera5,sera7,CDC_97}.
Where more than one antisera assay has been performed,
the calculated distances are averaged.
\label{table1}}
\end{sidewaystable}

}

\clearpage

\clearpage

\clearpage

\begin{figure}[t]
\begin{center}
\epsfig{file=figure3.eps,height=3in,clip=,angle=0}
\end{center}
\caption{Vaccine efficacy as observed in epidemiological studies
for influenza-like illness as a function of $p_{\rm sequence}$ (see
eq.\ \ref{eq_sequence}).
Also shown is a linear least squares fit
to the data (long dashed, $R^2 = 0.57$).
The epidemiological data shown in this figure  are the same 
as in figure \ref{fig:epitope}.  Only the definition of the 
$x$-axis is different.
\hspace{3in}
\label{fig:sequence}}
\end{figure}

\clearpage

\clearpage

\end{document}